\documentclass[square]{ws-procs11x85}
\usepackage{multicol,float} 

\begin{document}

\title{Gamma-ray emission from massive star forming regions}

\author{A. T. ARAUDO$^*$ and G. E. ROMERO$^{**}$}

\address{Instituto Argentino de
Radioastronom\'{\i}a, C.C.5, (1894) Villa Elisa, Buenos Aires,
Argentina \\
Facultad de Cs. Astron\'omicas y Geof\'{\i}sicas,
Universidad Nacional de La Plata, Paseo del Bosque, 1900 La Plata,
Argentina \\
$^*$E-mail: aaraudo@fcaglp.unlp.edu.ar\\
$^{**}$E-mail: romero@fcaglp.unlp.edu.ar}

\author{V. BOSCH-RAMON}

\address{Max Planck Institut f\"ur Kernphysik, Saupfercheckweg
1, Heidelberg 69117, Germany \\
E-mail: Valenti.Bosch-Ramon@mpi-hd.mpg.de}

\author{J. M. PAREDES}

\address{Departament d'Astronomia i
Meteorologia, Universitat de Barcelona, Mart\'{\i} i Franqu\`es 1,
08028, Barcelona, Spain \\
E-mail: jmparedes@ub.edu}

\begin{abstract}
Recent radio observations support a picture for star formation where there is
accretion of matter onto a central protostar with the ejection of
molecular outflows that can affect the surrounding medium. The impact
of a supersonic outflow on the ambient gas can produce a strong shock
that could accelerate particles up to relativistic energies.
A strong evidence of this has been the detection of non-thermal radio 
emission coming from the jet termination region of some young massive stars.
In the present contribution, we study the
possible high-energy emission due to the interaction of relativistic
particles, electrons and protons, with the magnetic, photon and matter fields 
inside a giant molecular
cloud. Electrons lose energy via relativistic Bremsstrahlung, 
synchrotron radiation and inverse Compton interactions, and protons cool
mainly through inelastic collisions with atoms in the cloud. 
We conclude that some massive young
stellar objects might be detectable at gamma-rays by next generation
instruments, both satellite-borne and ground based.
\end{abstract}


\bodymatter

\begin{multicols}{2}

\section{Introduction}\label{aba:sec1}

The mechanism of formation of massive stars ($M > 8M_{\odot}$) remains 
one of the open
questions in astrophysics. It is known that these stars
originate inside giant and massive molecular clouds but the sequence of
processes that take place during the formation of the star are mostly
unknown. It has been suggested, for example, that the coalescence of
various protostars in the same cloud can lead to the emergence of a
massive star \cite{Bonnell98}. Massive stars
appear in massive stellar associations where cloud fragmentation seems
to be common. Alternatively, a massive star could form by the collapse
of the core of a molecular cloud, with associated episodes of mass
accretion and ejection, as observed in low-mass stars \cite{Shu87}.
In such a case, the effects of jets propagating
through the medium that surrounds the protostar should be detectable.

Until now, the formation of stars has been mostly associated with   
thermal radio and X-ray emission. However,
non-thermal radio emission has been detected in some massive star 
forming regions. This is a clear evidence that efficient particle 
acceleration is occurring there, which may have as well a radiative 
outcome at energies much higher than radio ones.

In the present contribution, and based on recent multiwavelength 
observations, as well as reasonable physical assumptions, 
we show that massive protostars
could produce a significant amount of radiation in the gamma-ray domain, 
because of the dense and rich medium in which they are formed.

\section{Non-thermal radio sources}

In the last years, synchrotron radiation have been observed from some 
regions where massive stars form.  This
emission is associated with outflows emanating from a central protostar. 
In what follows, we briefly describe some of
these non-thermal sources that could be potential emitters of gamma-rays. 

\subsection{IRAS~16547$-$4247}

 The triple radio source associated with the protostar IRAS~16547$-$4247  
is one of the best candidates to produce
gamma-rays. This system is located  within a very dense region (i.e. 
densities $n\approx 5\times 10^5$~cm$^{-3}$) of
a giant molecular cloud located at a distance of 2.9~kpc. The luminosity
of the IRAS source is $L = 6.2\times 10^4
L_{\odot}$, possibly being the most luminous known YSO associated with 
collimated thermal jets. 

ATCA and VLA observations \cite{Garay03, Rodriguez05} have shown that 
the southern lobe of 
this system, of size $\sim 10^{16}$~cm, has a clear non-thermal spectrum,
with an index $\alpha \sim 0.6$ ($S_{\nu} \propto \nu^{-\alpha}$). 
The specific flux of this source is 2~mJy at
14.9~GHz and the estimated magnetic field is  $B \sim 2\times10^{-3}$~G 
\cite{Araudo07}.

\subsection{Serpens}

The Serpens molecular cloud is located at a distance of $\sim$ 300 pc. 
One of the
two central dense cores of this cloud is a triple radio source, composed 
by a central protostar (IRAS~18273$-$0113) and two lobes. The northwest (NW) 
hot-spot is connected with the central source by a highly collimated 
thermal jet, 
whereas the southeasth (SE) is separated and broken into two clumps.
The luminosity of the source IRAS~18273$-$0113 is 
 $L \sim 300 L_{\odot}$ and the particle density at the center of the 
molecular cloud is $n_0 \sim 10^5$ cm$^{-3}$.

The observed radio emission \cite{Rodriguez89, Curiel93}
detected from the central and NW sources has a spectral index 
$\alpha \approx -0.15$ and $\alpha \approx 0.05$, respectively. 
This emission, of a luminosity $\sim 2-3$~mJy, is produced via thermal 
Bremsstrahlung. 
However, the radiation produced in the SE lobe seems to be
non-thermal ($\alpha = 0.3$), likely produced via synchrotron emission.
The specific flux of this source is $2-5$~mJy. The equipartition magnetic 
field estimated in the SE lobe is $B_{\rm{equip}} \sim 10^{-3}$~G 
\cite{Rodriguez89}. 

\subsection{HH 80-81}

The famous Herbig-Haro objects called HH~80$-$81 are the south component of
a system of radio sources, located in the Sagitarius
region, at a distance of 1.7 kpc.  
The central source has been identified with the luminous 
($L = 1.7\times 10^4 L_{\odot}$) protostar IRAS~18162$-$2048.  HH~80 North 
is the northern counterpart of HH~80$-$81.
The velocity of the jet has been estimated as $v \sim 700$~km~s$^{-1}$, 
allowing to derive a dynamical age for the system similar to 4000~yr.

Radio observations carried out with the VLA instrument \cite{Marti93}
showed that the central source has a spectral index 
$\alpha \sim 0.1$, typical of free-free emission, whereas
HH~80$-$81 and HH~80 North are likely non-thermal sources, with spectral index 
$\alpha \sim 0.3$. The specific flux measured at a 
frecuency of 5 GHz is
$F_{\nu}\sim 1-2$~mJy and $F_{\nu}\approx 2.4$~mJy for the sources
HH~80$-$81 and HH~80 North, respectively. At this frecuency, the angular
size of the north and the south components are $\sim 6^{\prime\prime}$.

In addition, HH~80$-$81 system is a powerful source of thermal X-rays with 
a luminosity $L_X \sim 4.3\times
10^{31}$~erg~s$^{-1}$ \cite{Pravdo04}.

\subsection{W3(OH)}

Other interesting source to study is the system composed by an H$_2$O maser
complex  and the Turner-Welch (TW) source in the W3 region \cite{Wilner99}. The
central source of this system is a luminous  ($L \sim 10^5 L_{\odot}$) YSO and
the mean density of cool particles is  $n \sim 4\times10^4$~cm$^{-3}$. The
distance to W3(OH) is 2.2~kpc.

Continuum radio observations \cite{Wilner99} show the presence of a 
sinuous double-sided
jet, emanating from the TW source. The observed radio flux, from 1.6 to
15 GHz, is in the range 2.5-0.75 mJy, and the spectral index of the
observed emission is clearly non-thermal: $\alpha = 0.6$. 
The inhomogeneous synchrotron model proposed by Reid et al. in Ref. 
\cite{Reid95} 
predicts for the emitting jet a density of relativistic electrons 
$n_e(\gamma, r) = 0.068 \gamma^{-2}(r/r_0)^{-1.6}$~cm$^{-3}$ and a 
magnetic field 
$B(\gamma, r) = 0.01 (r/r_0)^{-0.8}$~G, 
where $r_0 = 6.6\times10^{15}$~cm, and $r$ and $\gamma$ are the distance 
to the jet origin and the electron Lorentz factor, respectively. Unlike 
in previous cases, here the non-thermal 
radio emission comes from the jet and not from its termination region. 
Non-thermal jets associated to a YSO are uncommon.  
 
\section{Acceleration of particles and losses}

The non-thermal radio emission observed in some massive star forming
regions is interpreted as synchrotron radiation produced by the
interaction of relativistic electrons with the magnetic field present
in the cloud, being typically $B_{\rm{cloud}} \sim 10^{-3}\;\rm{G}$ 
\cite{Crutcher99}. 
These particles
could be accelerated at a shock, formed in the point where the jet
terminates, via  diffusive shock acceleration \cite{Drury83}.  
The acceleration efficiency, characterized by $\eta$, is related to 
the velocity 
of the shock. Using the values of the velocities
given for the sources
described in the previous section, and assuming Bohm diffusion, 
values for $\eta$ of $\sim 10^{-6}-10^{-5}$ are obtained.

Particles accelerated up to relativistic energies interact with the 
different fields present in the medium. As
noted at the beginning of this section, electrons radiate synchrotron 
emission under the 
ambient magnetic field $B$. In addition, particles,
electrons and protons, can also interact with the cold matter in 
the jet termination region (via relativistic
Bremsstrahlung the leptons, and proton-nuclei collisions the protons). 
In addition, electrons interact with the
background field of IR photons of the protostar, of energy density 
$u_{\rm{ph}}$, via inverse Compton (IC) scattering.

\begin{figure}[H] 
\centerline{\psfig{file=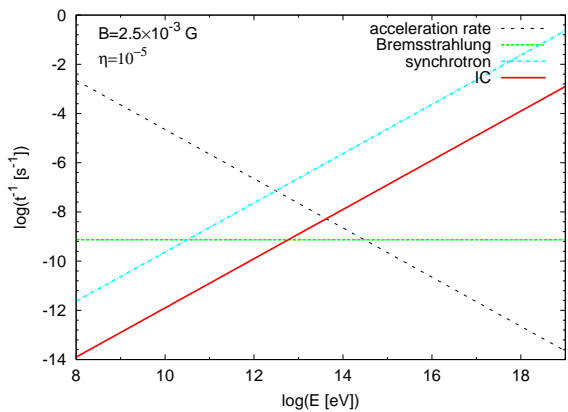,width=5.5cm, angle=270}}
\caption{Energy loss and acceleration rates for electrons in the 
IRAS~16547$-$4247 southern lobe.}
\label{fig1}
\end{figure}

Using the following parameter values: 
$n=5\times10^{5}$~cm$^{-3}$; $B=2.5\times10^{-3}$~G; 
and $u_{\rm ph}=3.2\times10^{-9}$~erg~cm$^{-3}$ given for 
IRAS~16547$-$4247, \cite{Garay03, Araudo07},
we  estimate the cooling time of the main leptonic processes in this 
scenario. As seen in Fig. \ref{fig1},
relativistic Bremsstrahlung losses are dominant up to $\sim 10$~GeV. 
In addition, it is possible to see from
this figure that the maximum energy achieved by electrons is 
$E_e^{\rm{max}} \sim 1$~TeV and is determined by
synchrotron losses, being IC losses negligible.

Protons can be accelerated by the shock in the same way as electrons and 
interact with cold particles present in the
cloud. The maximum energy achieved by protons is higher: 
$E_p^{\rm{max}} \sim 10$~TeV. In $pp$ interactions, besides
$\gamma$-rays, secondary electron-positron pairs are produced. 
These secondary particles will radiate by the same
mechanisms as primary electrons (i.e.  synchrotron radiation, IC 
scattering and relativistic Bremsstrahlung).

\section{High-energy emission}

\begin{figure*}[] 
\centerline{\psfig{file=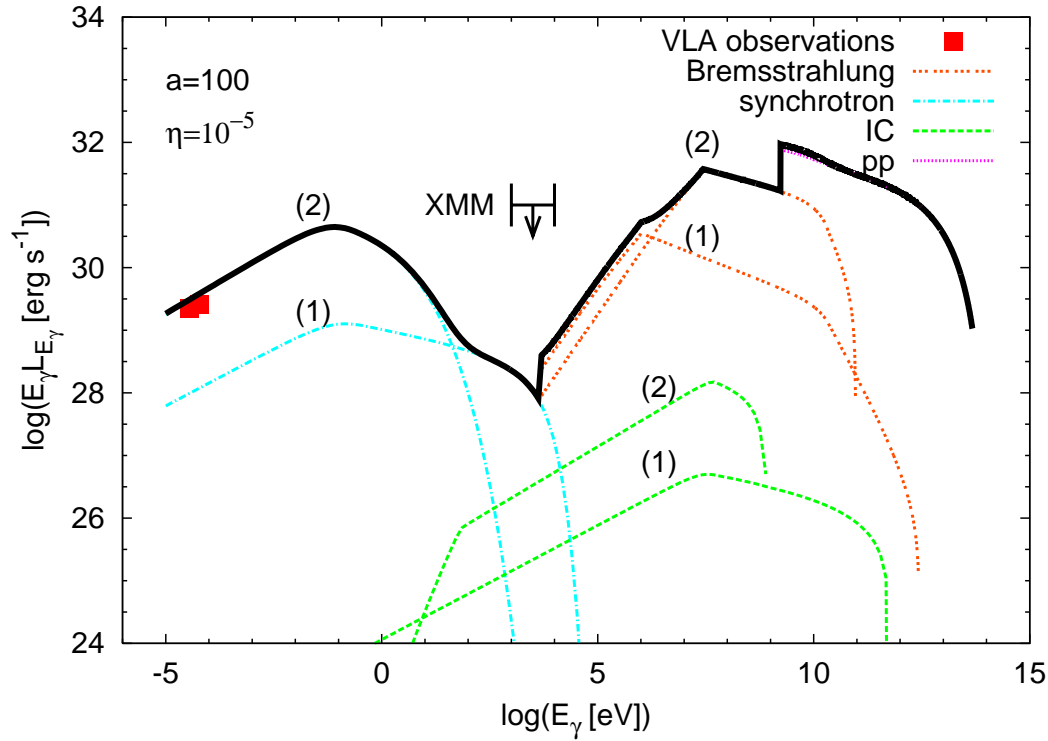,width=10cm, angle=270}}
\caption{Spectral energy distribution for the south lobe of the
YSO embedded in the source IRAS~16547$-$4247. The radiative contribution of
secondary pairs $(2)$ produced via $\pi^\pm$-decay is shown along with 
the contribution of primary electrons $(1)$.}
\label{fig2}
\end{figure*}

In order to calculate the non-thermal spectral energy distribution (SED) of
a massive protostar, the magnetic 
field of the cloud and the 
distributions of relativistic particles, $n(E)$, are needed. 
To obtain the magnetic field $B$ and the normalization
constants of the particle distributions, we use the standard
equations given by Ginzburg \& Syrovatskii \cite{Ginzburg64} for the observed 
synchrotron flux and assuming equipatition between the magnetic 
and the relativistic particle (primary and secondary pairs and protons) 
energy densities:
\begin{equation}
\frac{B^2}{8\pi} = u_e + u_p + u_{e^{\pm}}, 
\label{equip}
\end{equation}
In \eref{equip}, the following relationships are implicit: 
$u_{\rm p}= a\,u_{\rm e}$ and  $u_{e^\pm}= f\,u_{\rm p}$. 
In the first relation $a$ is a free 
parameter (here  fixed to 100, as it is the case for cosmic rays) and in
the second relation $f$ can be 
estimated using the average ratio of the number of secondary pairs to 
$\pi^0$-decay photons in Ref. \cite{Kelner06}.
In Fig. \ref{fig2}, we show the computed broadband 
SED for the parameters of the source 
IRAS~16547$-$4247. 

\section{Discussion}

In this work we show that, if the source is located at few kpc, 
the high-energy emission may be
detected by GLAST and even by forthcoming Cherenkov telescope arrays after 
long enough exposure.  This opens a new
window to the study of star formation and related processes. Also, 
determinations of the particle spectrum and its high-energy
for different sources with a variety of environmental conditions 
can shed light on the properties of
galactic, supersonic outflows, and on the  particle acceleration 
processes occurring at their termination points. 

Radio observations already demonstrate that relativistic electrons are 
produced in some sources. According to the
presence of non-thermal emission detected at cm-wavelengths and IR 
observations of the protostars emission we can
suggest several good candidates to be targeted by GLAST. These objects 
are IRAS~16547$-$4247 \cite{Araudo07}, 
HH~80$-$81  \cite{Marti93}, W3 \cite{Reid95, Wilner99} and the multiple 
radio source in Serpens \cite{Rodriguez89, Curiel93} 

To conclude, we emphasize that massive YSO with bipolar outflows and 
non-thermal radio emission can form a new
population of gamma-ray sources that could be unveiled by the next 
generation of $\gamma$-ray instrumentation.

\subsection*{Acknowledgements}

A.T.A. \& G.E.R. are supported by the Argentine Agencies CONICET 
(GRANT PIP 5375) and ANPCyT (GRANT PICT 03-13291 BID 1728/OC-AR).
V.B-R., and J.M.P  acknowledge support by DGI
of MEC under grant AYA2007-6803407171-C03-01, as well as partial support by
the European Regional Development Fund (ERDF/FEDER).
V.B-R. gratefully acknowledges support from the Alexander von Humboldt
Foundation.

\bibliographystyle{ws-procs11x85}
\bibliography{ws-pro-sample}

\begin{thebibliography}{9}

\bibitem{Bonnell98} I.~A. Bonnell, M.~R. Bate \& H. Zinnecker
  {\em MNRAS} {\bf 298}, 93 (1998).

\bibitem{Shu87} F.~H. Shu, F.~C. Adams \& S. Lizano
  {\em ARA\&A} {\bf 25}, 23 (1987).

\bibitem{Garay03} G. Garay, K. Brooks, D. Mardones \& R.~P. Norris
  {\em ApJ} {\bf 537}, 739 (2003).

\bibitem{Rodriguez05} L.~F. Rodri{\'\i}guez, G. Garay, K.~J. Brooks \&
D. Mardones
  {\em ApJ} {\bf 626}, 953 (2005).

\bibitem{Araudo07} A.~T. Araudo, G.~E. Romero, V. Bosch-Ramon \&  
J.~M. Paredes
  {\em A\&A} {\bf 476}, 1289A (2007).

\bibitem{Rodriguez89} L.~F. Rodr{\'\i}guez, S. Curriel, J.~M. Mor\'an, 
I.~F. Mirabel, M. Roth \& G. Garay
  {\em ApJ} {\bf 346}, L85 (1989).

\bibitem{Curiel93} S. Curriel, L.~F. Rodr{\'\i}guez, J.~M. Mor\'an 
\& J. Cant\'o
  {\em ApJ} {\bf 415}, 191 (1993).

\bibitem{Marti93} J. Mart{\'\i}, L.~F. Rodr{\'\i}guez, \& B. Reipurth
  {\em ApJ} {\bf 449}, 184 (1993).

\bibitem{Pravdo04} S.~H. Pravdo, Y. Tsuboi \& Y. Maeda
  {\em ApJ} {\bf 605}, 259 (2004).

\bibitem{Wilner99} D.~J. Wilner, M.~J. Reid, \& K.~M. Menten
  {\em ApJ} {\bf 513}, 775 (1999).

\bibitem{Reid95} M.~J. Reid, A.~L. Argon, K.~M. Menten, \& J.~M. Moran
  {\em ApJ} {\bf 443}, 238 (1995).

\bibitem{Crutcher99} R.~M. Crutcher
  {\em ApJ} {\bf 520}, 706 (1999).

\bibitem{Drury83} L.~O'C. Drury
  {\em Rep. Prog. Phys.} {\bf 46}, 973 (1983).


\bibitem{Ginzburg64} V.~L. Ginzburg \& S.~I. Syrovatskii {\it The Origin of 
Cosmic Rays},
  (Pergamon Press, New York, 1964).

\bibitem{Kelner06} S.~R. Kelner, F.~A. Aharonian, \&  V.~V. Vugayov
  {\em Phys. Rev. D} {\bf 74}, 034018 (2006).

\end{thebibliography}

\end{multicols}
\end{document}